\documentstyle{mn}
\newcommand{\reference}{\bibitem}

\def\beq{\begin{equation}}
\def\eeq{\end{equation}}
\def\bey{\begin{eqnarray}}
\def\eey{\end{eqnarray}}
\def\beqarray{\begin{eqnarray}}
\def\eeqarray{\end{eqnarray}}
\def\apj{ApJ}
\def\Dls{D_{\rm ls}}
\def\Ds{D_{\rm s}}
\def\Dl{D_{\rm l}}
\def\Mbh{M_{\rm bh}}
\def\Sig{\Sigma}
\def\Scr{\Sig_{\rm cr}}
\def\mcr{m_{\rm cr}}
\def\rcr{{R}_{\rm cr}}
\def\rc{r_{\rm c}}

\def\rs{r_{\rm s}}

\def\kpc{\,{\rm {kpc}}}
\def\pc{\,{\rm {pc}}}
\def\kpch{\,{h^{-1}{\rm kpc}}}

\def\kms{\,{\rm {km\, s^{-1}}}}
\def\kc{\kappa_{\rm c}}
\def\rcci{r_{\rm crit,1}}
\def\rcai{r_{\rm caust,1}}
\def\rccii{r_{\rm crit,2}}
\def\rcaii{r_{\rm caust,2}}
\def\rcciii{r_{\rm crit,3}}
\def\rcaiii{r_{\rm caust,3}}
\def\gbh{\gamma_{\rm BH}}
\def\delr{\Delta r}
\def\delrs{\Delta r_{\rm s}}
\def\muc{\mu_{\rm c}}

\input epsf

\title[]{The Influence of Central Black Holes on Gravitational Lenses}

\author[Mao, Witt \& Koopmans]
{Shude Mao,$^1$\thanks{e-mails: smao@jb.man.ac.uk, hwitt@aip.de,leon@jb.man.ac.uk}
 Hans J. Witt$^2$ and Leon V.E. Koopmans$^1$ \\
$^1$University of Manchester, Jodrell Bank Observatory, 
Macclesfield, Cheshire SK11 9DL, UK \\
$^2$Astrophysikalisches Institut Potsdam, An der Sternwarte 16, 
14482 Potsdam, Germany}

\date{Accepted ........
      Received .......;
      in original form .......}

\pubyear{2000}

\begin{document}
\maketitle
\begin{abstract}
Recent observations indicate that many if not all galaxies host
massive central black holes. In this paper we explore the influence of
black holes on the lensing properties. We model the lens as an
isothermal ellipsoid with a finite core radius plus a central black
hole. We show that the presence of the black hole substantially
changes the critical curves and caustics.  If the black hole mass is
above a critical value, then it will completely suppress the central
images for all source positions.  Realistic central black holes likely
have masses below this critical value. Even in such sub-critical
cases, the black hole can suppress the central image when the source
is inside a zone of influence, which depends on the core radius and
black hole mass.  In the sub-critical cases, an additional image may
be created by the black hole in some regions, which for some radio
lenses may be detectable with high-resolution and large dynamic-range
VLBI maps.  The presence of central black holes should also be taken
into account when one constrains the core radius from the lack of
central images in gravitational lenses.
\end{abstract}


\section{INTRODUCTION}

Recent observations suggest that many if not all nearby galaxies host
massive central black holes (e.g.  Kormendy \& Richstone 1995;
Magorrian et al. 1998). The existence of such central black holes can
be accommodated by hierarchical structure formation theories
(e.g. Silk \& Rees 1998; Kauffmann \& Haehnelt 2000). The effects of
central black holes on gravitational lensing have not been studied in
detail, although it is commonly believed that the central singularity
can suppress the central image. This is important because lensing
theories predict that any {\it non-singular} lens should have an odd
number of images while the observed lenses always show an even number
of images (Burke 1981; see also Schneider et al. 1992 for a general
review on gravitational lensing); the only possible exception is APM
08279+5255 (Ibata et al. 2000). The lack of central images could
indicate either that the potential is singular or that the central
surface density in lenses is so high that the central image is highly
demagnified and thus unobservable in current surveys (Narayan et
al. 1984).  The purpose of this paper is to clarify, through simple
examples, the effects of central black holes on lensing properties. We
show that the presence of black holes can not only suppress but also
create additional images. These additional images may be observable in
some cases. Central black holes also introduce somewhat unusual
critical curves and caustic structures. The outline of this paper is
as follows. In section 2, we first outline the lensing basics and
derive some analytical results for the critical curves, caustics and
image properties for an isothermal sphere plus a black hole. We then
generalize to the case of an isothermal ellipsoid plus a black hole.
In section 3, we discuss the implications of black holes on
gravitational lenses, including the constraints on the core radius.

\section{EFFECTS OF CENTRAL BLACK HOLES ON LENSING}

We model a lensing galaxy as an isothermal ellipsoid plus a central
black hole. The isothermal ellipsoid model is not only analytically
tractable but also consistent with models of individual lenses, lens
statistics, stellar dynamics, and X-ray galaxies (e.g.\ Fabbiano 1989;
Maoz \& Rix 1993; Kochanek 1995, 1996b; Grogin \& Narayan 1996; Rix et
al.\ 1997). The surface density distribution of an isothermal
ellipsoid is given by
\beq \label{eq:kappa}
  {\Sigma \over \Sigma_{\rm cr}} = {1 \over 2 q}\,
    {1 \over \sqrt{x^2+y^2/q^2+\rc^2}}\ ,
\eeq
where
$\rc$ is the core radius, $q$ is the axial ratio, and $\Sigma_{\rm
cr}=c^2 \Ds/(4\pi G\Dl\Dls)$ is the critical surface density, $\Dl$,
$\Ds$ are angular diameter distances from the observer to the lens and
source, respectively, and $\Dls$ is the angular diameter distance from
the lens to the source. Notice that all the lengths ($x, y, \rc$) are
expressed in units of the critical radius, $\rcr$,
\beq \label{eq:units}
\rcr=\Dl b, \qquad b \equiv
4\pi \left({\sigma \over c}\right)^2\,{\Dls \over \Ds},
\eeq
where the critical angle $b$ is the angle extended by the critical
radius on the sky ($b \sim 0.2-3$ arcsecond for typical lens galaxies;
e.g. Jackson et al.  1998), and the velocity dispersion $\sigma$ in
eq. (\ref{eq:units}) is related to, but not identical to, the
observable line of sight velocity dispersion; we shall ignore this
minor complication in our analysis and simply treat it as a parameter
(see Keeton, Kochanek \& Seljak 1997 for further discussions).

The lensing properties of the isothermal ellipsoid have been given by
Kassiola \& Kovner (1993), Kormann, Schneider \& Bartelmann (1994),
and Keeton \& Kochanek (1998). The lens equation of an isothermal
ellipsoid plus a black hole is
\begin{eqnarray}
         u&=&x- {1 \over \sqrt{1-q^2}}\,\tan ^{-1}\left(\sqrt{1-q^2}\, x
	\over {\varphi + \rc}\right)-m {x \over r^2}\,, \nonumber\\ v &=& y-
	{1 \over \sqrt{1-q^2}}\,\tanh^{-1}\left(\sqrt{1-q^2}\, y \over
	{\varphi +q^2 \rc}\right)-m {y \over r^2}\,, \label{eq:lens}
\end{eqnarray}
where $\varphi^2=q^2(\rc^2+x^2)+y^2, r^2=x^2+y^2$, and the
dimensionless black hole mass is given by
\beq \label{eq:bhUnits}
m={\Mbh \over M_0}, \qquad M_0 \equiv { \pi\sigma^2\rcr \over G},
\eeq
with $\Mbh$ being the mass of the black hole. Physically, $M_0$ is the
mass of the galaxy contained within a cylinder with radius $\rcr$ and
hence $m_0$ is essentially the ratio of the black hole mass to the
mass of the galaxy in the inner parts (within $\rcr$). The
magnification ($\mu$) of a given image is given by
\beq \label{eq:jacobian}
	\mu^{-1} = {\partial{u} \over \partial{x}}
	{\partial{v} \over \partial{y}} - 
	{\partial{v} \over \partial{x}} {\partial{u} \over \partial{y}}.
\eeq
For definitiveness, we shall adopt a lens redshift of 0.5 and a source
redshift of 2. In the Einstein-de Sitter universe ($\Omega_0=1$ with
no cosmological constant), the angular and length units are given by
\beq 
	b=1^{''} \left({\sigma\over 250}\right)^2, \qquad
	\rcr=3.6\kpch \left({\sigma\over 250}\right)^2,
\eeq
where $h$ is the Hubble constant in units of $100\,{\rm
km\,s^{-1}\,Mpc^{-1}}$ and $\sigma$ is in units of ${\rm km\,s^{-1}}$.
The dimensionless black hole mass for the given lens and source
redshifts is
\beq \label{eq:bh}
	m=2.4 \times 10^{-3} h\left( {\sigma \over 250} \right)^{3.27},
\eeq
where we have used the tight correlation of the black hole mass with
velocity dispersion found by Ferrarese \& Merritt (2000), $\Mbh\approx
4\times 10^8 M_\odot (\sigma/250)^{5.27}$ (see Gebhardt et al. 2000
for a different scaling.)\footnote{We neglect any possible evolution
of the black hole mass with time.}

The lens equation (eq. {\ref{eq:lens}}) can be readily solved
numerically. In general there are several curves in the image plane
along which the magnification is infinite ($\mu^{-1}=0$).  These are
called ``critical curves'', and they map to ``caustics'' in the source
plane. Caustics mark discontinuities in the number of images, so in
order to determine the number of images produced by a lens model, it
is sufficient to examine the caustics. We shall consistently employ
this idea in this paper. In order to gain some analytical insights
into the influence of central black holes, we start with the spherical
case (i.e. $q=1$) in the next subsection.

\subsection{Spherical Isothermal Models with Black Holes}

For the spherical case ($q=1$), the surface density is given by
\beq
	{\Sigma \over \Scr} = {1 \over 2 \sqrt{r^2 + \rc^2}}.
\eeq
This profile has been studied quite extensively by Hinshaw \& Krauss (1987).

The lens equation for this profile in the presence of a black hole is
quite simple,
\begin{equation}\label{eq:lenseq}
	\rs = r-\alpha(r), \qquad \alpha(r)= {\sqrt{r^2+\rc^2}-\rc \over r} +
{m \over r},
\end{equation}
where $\rs, \rc$ are the source position and core radius in units of
the critical radius and $m$ the dimensionless mass of the central
black hole given in eq. (\ref{eq:bhUnits}).

\subsubsection{Critical Curves and Caustics}

The critical curves are given by the equation 
\beq 
	\mu^{-1} = {\rs \over r} {d\rs \over dr}=0.  
\eeq 
The solution satisfies either $\rs/r=0$ or $d\rs/dr=0$. One can verify
that the first condition ($\rs/r=0$) always yields one single critical
curve,
\beq\label{eq:poly1} 
	r^2={1 \over 2} \left[\,1-2\rc+2m+\sqrt{(1-2\rc)^2+4 m}\, \right], 
\eeq 
which maps to a
degenerate caustic point at the origin. The second condition
($d\rs/dr=0$) can be manipulated into a polynomial 
\begin{eqnarray}\label{eq:poly2} 
	r^6+(2 m - 2 \rc+\rc^2)r^4 & + & \nonumber \\ 
	(m^2-2m \rc + \rc^2+2 m \rc^2 - 2 \rc^3)r^2  & + & \nonumber \\  
          m^2 \rc^2 - 2 m \rc^3 & = & 0.  
\end{eqnarray}
This equation is cubic in $r^2$, and so can be solved
analytically. One finds that beyond a critical black hole mass, there
is no physical solution (i.e.  positive solution in $r^2$), while
below the critical mass, there are always two physical solutions.
The critical mass can be found by examining eq. (\ref{eq:poly2}); some
algebra yields the critical mass 
\beq \label{eq:crit} 	
	\mcr = \rc (1 +\rc) - 3 \left({\rc^2 \over 2}\right)^{2/3}.  
\eeq 
For a given core radius, the central images are always suppressed by
the black hole if its mass exceeds the critical value, $\mcr \approx
\rc$ when $\rc\ll1$. This can be understood intuitively as follows:
when the mass is comparable to $\rc$, the mass of the black hole is
comparable to the galaxy mass enclosed by the core radius, so the
black hole makes the deflection angle nearly flat at $r \sim \rc$,
similar to the singular isothermal case, for which we have only two
images (cf. Fig. 2). Notice, however, that even when the black hole
mass is below the critical value, there is a region in the source
plane where the central image is suppressed (see below).

We can obtain approximate but more illuminating solutions for the
positions of critical curves and caustics when $\rc \ll 1$ and
$m\ll\mcr \approx \rc$. The first assumption ($\rc \ll 1$) is
supported by the lack of central images in lenses (e.g.  Wallington \&
Narayan 1993; Kochanek 1996a; see also \S2.1.2), while the second
condition is valid for most galaxies as can be seen from
eq. (\ref{eq:bh}), and if $\rc \ga 0.002$ ($\sim 10$ pc for
$\sigma=250\kms$).  Under these two assumptions, the inner most
critical curve satisfies $r \ll \rc$ (justified below), the lens
equation (eq. \ref{eq:lenseq}) simplifies to
\beq \label{eq:lenseqsimple}
	\rs=(1-\kc)r-{m\over r},
\eeq
where $\kc\equiv {1/(2 \rc)}$ is the central surface density of the
galaxy, in the absence of the black hole. From this simplified lens
equation, one can readily show that the innermost critical curve forms
where $d\rs/d r=0$, i.e.
\beq\label{eq:rcci}
	\rcci = \sqrt{m \over {\kc -1}}.
\eeq
The second assumption (i.e. $m\ll\mcr$) ensures that our condition, $r
\ll \rc$, is satisfied.  The corresponding caustic in the source plane
is
\beq\label{eq:rcai}
	\rcai = -2\, \sqrt{m (\kc -1)} = {-2\, m\, \rcci^{-1}}.
\eeq
The middle critical curve in general does not satisfy the condition $r
\ll \rc$. To derive its approximate position, we simplify
eq.\,(\ref{eq:poly2}) by setting $m=0$, after which we are left with a
quadratic equation in $r^2$. One finds that, to the leading order of
$\rc$, the positions of the critical curves and caustics are
\beq\label{eq:rccii}
	\rccii \approx \sqrt{\rc}, ~~
	\rcaii \approx 1-2\sqrt{\rc} = 1-2\,\rccii.
\eeq
 From eq.\,(\ref{eq:poly1}) one finds that the outer most critical curve
and its corresponding caustic are at
\beq\label{eq:rcciii}
	\rcciii \approx 1-\rc, ~~\rcaiii=0.
\eeq
For a source located outside $\rcaii$ there are only two images. When a
source is between $\rcai$ and $\rcaii$, there are four images; compared
with the case without a black hole, one additional image has been created very
close to the black hole. For a source inside $\rcai$, there are only 
two images, the central image that would be present without a black hole
has been suppressed. The radius $\rcai$ can therefore be regarded as a
measure of the zone of influence of the black hole: for any source 
$\rs<\rcai \approx (2m/\rc)^{1/2}$, the central image is destroyed.
Notice that this zone of influence scales as $\propto m^{1/2}$, identical to
the scaling of the Einstein radius of an isolated black hole with mass.

\begin{figure}
\epsfxsize=\hsize
\begin{center}
\epsfbox{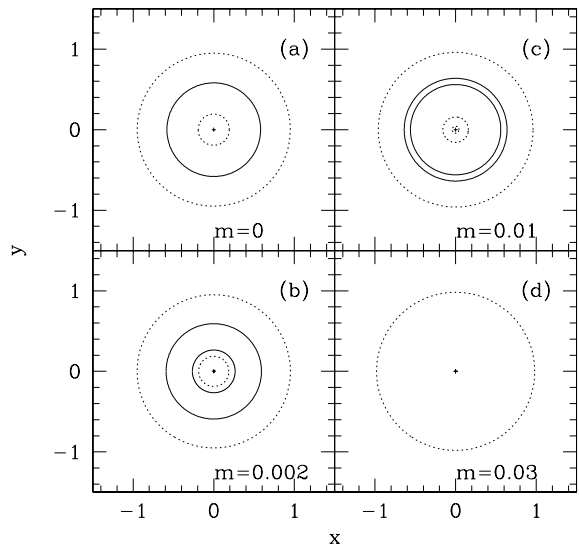}
\end{center}
\caption{
The critical curves (dotted lines) and caustics (solid lines) for a
spherical isothermal sphere with a black hole.  The core radius is
$\rc=0.05$.  All the lengths are in units of the critical radius
(eq. \ref{eq:units}). Each panel is labeled by the dimensionless black
hole mass ($m$), given in eq. (\ref{eq:bhUnits}), which is essentially
the ratio of the black hole mass to the mass of the galaxy within the
critical radius.  The critical mass for this case is $m \approx
0.0175$.  Notice that there is a degenerate caustic point at the
origin, indicated by a small plus sign. There is a tiny critical curve
around the origin for $m=0.002$, but is too small to see clearly.}
\end{figure}

Fig. 1 illustrates the evolution of critical curves and caustics with
the mass of the black hole increasing from zero to 0.03.  The core
radius is taken to be 0.05, a somewhat arbitrary value which satisfies
the upper limit inferred from the lack of central images (e.g.
Wallington \& Narayan 1993; Kochanek 1996a).  When there is no black
hole at the center (Fig. 1a) the critical curves are two circles, the
outer critical curve maps to a degenerate caustic point at the origin,
while the inner critical curve maps to the solid circle. Fig. 1b shows
the configuration when we include a black hole mass with $m=0.002$, a
value expected for a galaxy with $\sigma \approx 250{\rm\,km\,s^{-1}}$
(cf. eq. \ref{eq:bh}). In addition to the two critical curves found
when $m=0$, an additional tiny critical curve appears around the
origin (too small to be seen), which maps to the inner solid
circle. The positions of the three critical curves and their
corresponding caustics are well approximated by
eqs. (\ref{eq:rcci}-\ref{eq:rcciii}). For a source moving from
infinity to the origin, it would first have two images, then four
images as the source moves across the first caustic. Compared with the
case without a black hole, one additional image has been created very
close to the black hole. The number of images decreases to 2 again as
the source moves across the second caustic.  Notice that even in this
case where the black hole mass is substantially below the critical
mass $m=\mcr \approx 0.0175$, the zone of influence determined by
$\rcai \approx 0.28$ is substantial. Within the zone of influence, the
central image is completely suppressed. When the black hole mass is
further increased to $m=0.01$ (Fig. 1c), the inner-most critical curve
becomes larger while the middle critical curve shrinks, their
corresponding caustics also approach each other. When the black hole
mass is increased to the critical mass $\mcr$ (eq. \ref{eq:crit}), the
two critical curves merge into a single curve, so do their
corresponding caustics. When the black hole mass is increased above
the critical value (an example is shown for $m=0.03$ in Fig. 1d), one
is left with a single critical curve and a degenerate caustic point at
the center. In this case, there are only two images no matter where
the source is. Notice, however, that the black hole mass for the
panels 1c and 1d are somewhat unrealistic (cf. eq. \ref{eq:bh}.)

\begin{figure}
\epsfxsize=\hsize
\begin{center}
\epsfbox{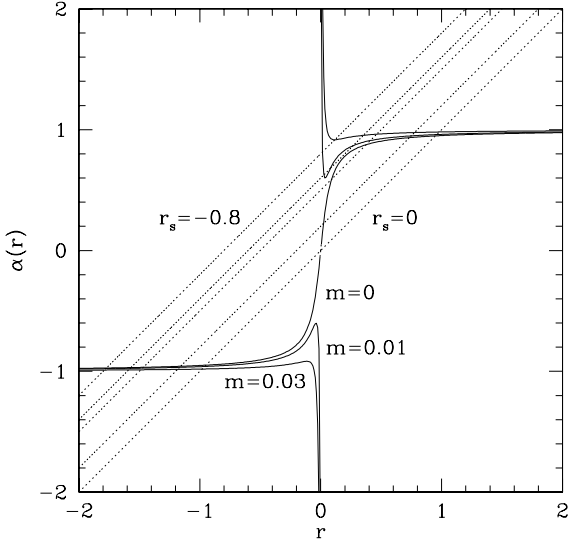}
\end{center}
\caption{
Deflection angle vs. the radius, shown for three black hole masses,
$m=0, 0.01$ and 0.03 (from bottom to top at positive $r$).  The
parameter $m$ is the ratio of the black hole mass to the mass of the
galaxy contained within the cylinder of radius equal to the critical
radius. The five dotted straight lines indicate $\rs=-0.8, -0.6, -0.5,
-0.2$ and 0 (from top to bottom). For each source position, the image
positions are given by the intercepts of the deflection angle curve
with the straight line corresponding to the source position. Notice
how the image number and position change as the source position
changes.}
\end{figure}

The change of image numbers may first appear puzzling, but can be
easily understood in the conventional diagram of the deflection angle,
$\alpha(r)$, vs. radius, shown in Fig. 2. Three examples are shown for
$\rc=0.05$ corresponding to three black hole masses, $m=0, 0.01$, and
0.03. For each source position, the image positions are given by the
the intercepts of the deflection angle curve with the straight line
$f(r)=r-\rs$.  For example, it is easy to see why for $m=0.01$, there
are three critical curves. It is clear from this figure that the
central black hole does not perturb the two outer images very much, as
expected, it only strongly perturbs the central images.

\subsubsection{Image Positions and Magnifications}

In this subsection, we will obtain approximate image positions and
magnifications when there are four images, i.e., when the source
position is between $\rcai$ and $\rcaii$. We are particularly
interested in the case when the source is close to $\rcai$. For such
cases, there are usually two bright outer images and two faint central
images; one of the fainter central image is created by the black
hole. We would like to address the question whether such images are
observable.

Using the resultant technique as discussed in Witt \& Mao (1995), one
can verify that the four image magnifications must satisfy an exact
relation
\beq
	\mu_1+\mu_2+\mu_3+\mu_4=2.
\eeq
Note that all the magnifications include their parities (signs).

The brightest image has positive parity and is located in regions
where $r \gg \rc$. For this image, the lens equation can be simplified
into $r^2 -\rs r +\rc -r = 0$, where we have only retained terms to
the order of $(\rc/r)^2$. From this equation, we can obtain the
position and magnification of the brightest image
\beq \label{mu1}
	r_1 \approx 1 + \rs -\rc, \quad \mbox{and} \quad
\mu_1^{-1} \approx \frac{\rs}{r_1^3} (r_1^2 + m-\rc) \approx \frac{\rs}{r_1}.
\eeq
Similarly, we can obtain the position for the bright outer image that 
has negative parity
\beq \label{mu2}
	r_2 \approx -1 + \rs +\rc, \quad \mbox{and} \quad
\mu_2^{-1} \approx \frac{\rs}{r_2}.
\eeq
For singular isothermal spheres ($\rc=0$), the magnifications and
positions are exact for these two images, and they satisfy
$\mu_1+\mu_2=2$.

For the two central images, we can estimate their positions and
magnifications if they satisfy $ r \ll \rc$. In this case, the lens
equation can be approximated by eq. (\ref{eq:lenseqsimple}) and the
image magnification is given by
\beq\label{eq:mui}
	\mu = \frac{1}{(1-\kc)^2 - m^2/r^4}.
\eeq
For very small $m$ we can obtain for the positions and magnifications
for the third and fourth image
\beq \label{image3}
	r_3 \approx - \frac{\rs}{\kc-1} \quad \mbox{and} \quad
\mu_3 \approx \muc, ~~ \muc \equiv \frac{1}{(\kc-1)^2}.
\eeq
\beq \label{image4}
	r_4 \approx -\frac{m}{\rs} \quad \mbox{and} \quad \mu_4^{-1} \approx
-\frac{\rs}{r_4^3} (\rs r_4 + 2m) = -\frac{\rs^4}{m^2}.
\eeq
Note that the magnification for the third image $\mu_3 \sim \kc^{-2} =
4 \rc^2$, is independent of the source position when the black hole
mass is small.  Previous studies neglected the role of central black
holes and therefore the lack of central images immediately implies an
upper limit on the core radius. If the brightest image is magnified by
a factor of few, and no central image is seen with a dynamic range
(DR) of 100, then this would imply $\rc \la 0.1$, which is roughly the
published upper limits on the core radius from lenses (e.g. Wallington
\& Narayan 1993; Kochanek 1996a).  Hence the lack of central images in
lenses may mean either that the core radius is small (see \S 2.1.2) or
that the black hole mass is quite massive; we return to this point in
the discussion.

The fourth central image is usually the closest to the black hole and
is usually very faint. However, when the source lies very close to
$\rcai$, the second term in the denominator of eq. (\ref{eq:mui}),
which is the square of the shear ($\gbh$) from the black hole, can
become comparable to $(1-\kc)$, and then the magnification is of order
unity or higher.  The question is: how large is this region? To
estimate this, we combine eqs. (\ref{eq:rcci}) and (\ref{eq:mui}) and
find that the position of the image is related to its magnification by
\beq\label{eq:radmu}
	\left(\frac{r}{\rcci}\right)^4 = \left(1-\frac{\muc}{\mu}\right)^{-1},
\eeq
where $\muc$ is the magnification of an image formed well within the
core radius in the absence of the black hole and is given in
eq. (\ref{image3}).  A small displacement $\delr$ from the critical
curve is related to the magnification through
\beq
	\delr = \frac{\muc}{4\mu}\rcci.
\eeq
This displacement is related to a corresponding displacement of the
source in the source plane, through eq. (\ref{eq:lenseqsimple}). By
definition the first order term of a Taylor expansion of the lens
equation vanishes near the caustic. The second term, after some
algebra, gives
\beq
\delrs=-\frac{m}{\rcci^3}\times \delr^2=-\frac{\sqrt{m}\, \muc^{7/4}}
	{16 \mu^2}.
\eeq
The probability of the source in a multiply-imaged system, lying inside
this region (i.e. having a magnification $>\mu$), is therefore
\beq\label{eq:probmu}
P_{\rm BH}(>\mu)=\frac{2\rcai\delrs}{\rcaii^2}=\frac{m \muc^{3/2}}
	{4 \mu^2 (1-2\sqrt{\rc})^2}.
\eeq
For any realistic situation, this probability is exceedingly small
($\ll10^{-6}$); the inclusion of magnification bias and extended
source structure does not substantially enhance the probability of
this faint central image to have magnifications above unity. 

In case of high dynamic-range (DR) maps, however, central images with
$\mu\ll 1$ can still be detected. In such cases eq. (\ref{eq:probmu}) is no
longer valid and we have to find a new estimate for the probability. To
do this, we insert eq. (\ref{eq:radmu}) into the simplified lens
equation, eq. (\ref{eq:lenseqsimple}), and arrive at
\beq
  \rs = \frac{1}{2}\, \rcai \left[\left(1-\frac{\muc}{\mu}\right)^{-1/4}+
	 \left(1-\frac{\muc}{\mu}\right)^{+1/4}\right].
\eeq
For $\mu\rightarrow \infty$, the source position goes to $\rcai$, as
expected. In case $\mu=\mu_4=-10^{-3}$ (DR$\ga10^3$; the magnification
is negative because the parity of the central image created by the BH
is $-$1), the source position $\rs\approx 1.15 \,\rcai$, assuming
$\rc=0.05$ and $m=0.002$. The probability of observing this image is
then $\approx 2\pi \rcai(\rs-\rcai)/(\pi\rcaii^2)$, which is $\sim$8\%
for the above given values of $\rc$ and $m$. If $\rc\sim0.05$, the
currently known sample of 20--30 radio graviational lens systems could
therefore contain several systems with central images that are
observable, if the dynamic range in radio observations can be improved
by a factor of 5--10. To separate the two central images, which are
only a few milli-arcseconds apart, one obviously requires
high-resolution VLBI observations.

\subsection{Elliptical Isothermal Density Distribution With Black Holes}

\begin{figure}
\epsfxsize=\hsize
\begin{center}
\epsfbox{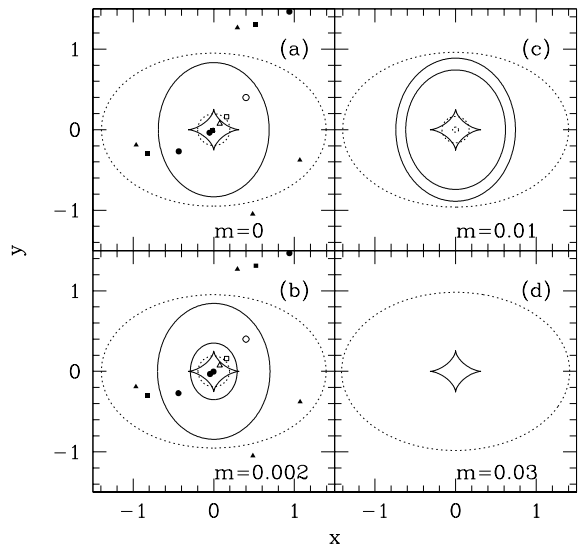}
\end{center}
\caption{
The critical curves (dotted) and caustics (solid) for an isothermal
ellipsoid plus a black hole. The core radius is taken to be $\rc=0.05$
and the axial ratio is 0.7.  All the lengths are in units of the
critical radius (eq. \ref{eq:units}), and the dimensionless black hole
mass ($m$) is given in units of eq. (\ref{eq:bhUnits}).  For the two
left panels the images are shown for three source positions, indicated
by an open triangle, square and circle; their corresponding image
positions are shown as filled triangles, squares and circles. For
Fig. 3a, a source outside the solid ellipse will have one image, and
then three images as the source crosses the elliptical caustic and
five images when the source crosses the diamond caustic. For bottom
left panel, as a source moves from infinity to the center, the number
of images changes according to $2\rightarrow 4 \rightarrow 2
\rightarrow 4$, with each caustic crossing changing the number of
images by 2. }
\end{figure}

The spherical case we studied in the previous subsection is idealized,
since galaxies nearly always show some ellipticity. In this section,
we will study the critical curves and caustics for a more realistic
elliptical density distribution plus a black hole.  The critical
curves can be found by solving eq. (\ref{eq:jacobian}) numerically and
the caustics are found using the lens mapping
(eq. \ref{eq:lens}). Fig. 3 shows the critical curves and caustics
structure as a function of black hole mass for an elliptical density
mass distribution with $q=0.7$ and again $\rc=0.05$. When there is no
black hole at the center (Fig. 3a), one sees the well known critical
curves and caustics.  When a source is inside the diamond caustic,
there are five images. When a source is outside the diamond but inside
the elliptical caustic, there are three images. When the source moves
further outside, there is only one single image. The image
configurations for three source positions are illustrated in the same
panel. Usually the central images are quite faint and difficult to
observe. Now as we include a black hole mass with $m=0.002$
(cf. eq. \ref{eq:bh}), an additional small critical curve appears very
close to center (too small to be seen), and this maps to the middle
elliptical caustic. A source outside the diamond but inside this
middle elliptical caustic has two images. This is analogous to
Fig. 1b.  When a source moves from infinity to the origin, the image
number changes according to $2\rightarrow 4 \rightarrow 2 \rightarrow
4$, with each caustic crossing either increases or decreases the
number of images by two.  When the black hole mass is further
increased to $m=0.01$ (Fig. 3c) the inner-most critical curve becomes
larger while the middle critical curve shrinks, their two
corresponding caustics also approach each other. When the black hole
mass is further increased to or above a critical value, $\approx 0.02$
(a value slightly larger than that in the spherical case), the two
inner critical curves and their corresponding caustics cancel each
other, and one is left with an elliptical critical curve and a diamond
caustic at the center (an example is shown for $m=0.03$); this caustic
separates the outer two image regions from the inner four image
regions. In either region, the central images have been suppressed by
the black hole. The behavior in the elliptical density distribution
case is qualitatively similar to the spherical case, the main
difference is that the ellipticity breaks the spherical symmetry and
changes the central degenerate caustic point in the spherical case
into the central diamond (compare Figs. 1 and 3.); the caustic change
due to ellipticity is similar to the case without central black holes
(see e.g. Schneider et al. 1992).

\section{Discussion}

As clearly demonstrated in Figs. 1--3, central black holes in
gravitational lenses introduces qualitatively new features in the
critical curves and caustics. For realistic black hole masses, the
black hole introduces a new region in the source plane that has four
images. However, this additional four-image region does not help to
resolve the problem of the apparent excess of quadruple lenses
relative to double lenses (King \& Browne 1996; Kochanek 1996b). The
reason is that these four image configurations have two very faint
images at the center and two bright images outside, which are very
different from the observed quadruple lenses where the images are
roughly on a circle from the lens center.  The image closest to the
black hole is the faintest image; for the case shown in Fig. 3b, this
image is about a factor of $10^{-5}$ times fainter than the brightest
one. As shown in section 2.1.1, there is only a negligible probability
that the central images can reach magnifications of order
unity. However, we find that there is a non-negligible probability
($\sim$8\% for $\rc=0.05$ and $m=0.002$) that the central image can
reach magnification $\ga 10^{-3}$, with the two distinct central
images separated by about $10^{-3}\,b$ ($\sim$ milli-arcseconds).
Such images can in principle be detected with future high dynamic
range and high resolution VLBI observations.  Note that this
probability obviously depends on the values of $m$ and $\rc$, and is
uncertain since for the latter only upper limits are available at
present.

The massive black holes also complicate the interpretation of core
radius inferred due to the lack of central images.  The central image
will be completely suppressed for all source positions if the black
hole mass exceeds the critical value as given in
eq. (\ref{eq:crit}). This critical mass is reached when the ratio of
the mass of the black hole to the galaxy mass in the inner parts (more
precisely, $M_0$ in eq. \ref{eq:bhUnits}) is approximately equal to
the core radius in units of the critical angle $b$.  For a galaxy with
velocity dispersion of $250\kms$, the mass of the central black hole
is $\sim 4\times 10^{8}M_\odot$, and $M_0 \sim 1.6 \times
10^{11}h^{-1}M_\odot$. The black hole is super-critical when $\rc \la
0.002$; in physical units, this implies an upper limit on the core
radius of about $\sim 10$ pc for typical lens configuration with $\rcr
\sim {\rm few} \kpc$. This upper limit is usually not reached except
for lens systems with favorable image geometry such as B1030+074
(Norbury et al. 2000, in preparation.) However, the presence of
central black holes can also suppress the central image even when it
is much below the critical value in some regions (see Figs. 1b and
3b). The zone of influence of the black hole is characterized by the
$\rcai$ (eq. \ref{eq:rcai}) in the spherical case: when the source is
inside the zone ($\rs<\rcai \approx (2m/\rc)^{1/2}$), the central
image is destroyed. For a given source position $\rs$, the central image
disappears when the black hole mass satisfies
\beq \label{eq:mlimit}
    m \ga \frac{\rc\rs^2}{2} \approx \frac{\rc}{2(\mu_1 - 1)^2},
\eeq 
where in the last step we have replaced the source position on the
right hand side by the magnification of the brightest image $\mu_1$
using eq. (\ref{mu1}).  For a galaxy with $\sigma=250\kms$, $m\approx
0.002$, and $\mu_1 \sim 4$, the central image will be destroyed when
$\rc <0.04$ ($\sim 100\pc$ for a typical critical radius of a few
kpc). More generally, the influence of black holes on the core radius
constraint has to be obtained from detailed modelling.
Eq. (\ref{eq:mlimit}) also highlights that the best lenses to search
for central images are systems with low magnifications.  In such
systems, the source is closer to outer caustic ($\rcaii$) and hence
more likely to be outside the zone of influence. As a result, the
central images are less likely to be suppressed by the black hole.
Unfortunately, these systems also have large flux ratios between the
two bright out images (cf. eqs. \ref{mu1}-\ref{mu2}). Since lens
surveys favor systems with large magnifications due to magnification
bias and systems with small flux ratio, it is not surprising that very
few lenses discovered so far show central images, perhaps as a result
of both small core radii and central black holes in gravitational
lenses.

\section*{Acknowledgments}

We thank Ian Browne for helpful discussions and Chuck Keeton for
insightful criticisms on a draft of the paper.

{}

\bsp
\label{lastpage}
\end{document}